\documentclass[preprint,showpacs,amsmath,amssymb]{revtex4}
\usepackage{epsfig}
\usepackage{rotating}
\usepackage{array}
\usepackage{epsfig}
\usepackage{graphicx}
\usepackage{epstopdf}
\usepackage{amssymb}
\usepackage{latexsym}

\newcommand{\be}{\begin{eqnarray}}
\newcommand{\bb}{\bibitem}
\newcommand{\ee}{\end{eqnarray}}

\newcommand{\fig}{\begin{figure}}
\newcommand{\ef}{\end{figure}}
\newcommand{\bc}{\begin{center}}
\newcommand{\ec}{\end{center}}

\newcommand{\bn}{\begin{enumerate}}
\newcommand{\en}{\end{enumerate}}

\newcommand{\bz}{\begin{itemize}}
\newcommand{\ez}{\end{itemize}}
\newcommand{\ct}{\centerline}
\newcommand{\ep}{\epsfig}
\newcommand{\cp}{\caption}

\newcommand{\ba}{\begin{array}} 
\newcommand{\ea}{\end{array}}

\newcommand{\bt}{\begin{tabular}}
\newcommand{\et}{\end{tabular}}

\newcommand{\mc}{\mathcal}

\newcommand{\bd}{\begin{displaymath}}
\newcommand{\ed}{\end{displaymath}}
\newcommand{\nn}{\nonumber}
\newcommand{\ben}{\begin{eqnarray*}}
\newcommand{\een}{\end{eqnarray*}}

\newcommand{\Ga}{\Gamma}

\newcommand{\bq}{\begin{quote}}
\newcommand{\eq}{\end{quote}}

\begin{document}
\def\bea{\begin{eqnarray}}
\def\eea{\end{eqnarray}}
\def\nn{\nonumber}
\newcommand{\snu}{\tilde \nu}
\newcommand{\sll}{\tilde{l}}
\newcommand{\asnu}{\bar{\tilde \nu}}
\newcommand{\stau}{\tilde \tau}
\newcommand{\dmsnu}{{\mbox{$\Delta m_{\tilde \nu}$}}}
\newcommand{\mt}{{\mbox{$\tilde m$}}}

\renewcommand\epsilon{\varepsilon}
\def\be{\begin{eqnarray}}
\def\ee{\end{eqnarray}}
\def\lla{\left\langle}
\def\rra{\right\rangle}
\def\za{\alpha}
\def\zb{\beta}
\def\lsim{\mathrel{\raise.3ex\hbox{$<$\kern-.75em\lower1ex\hbox{$\sim$}}} }
\def\gsim{\mathrel{\raise.3ex\hbox{$>$\kern-.75em\lower1ex\hbox{$\sim$}}} }
\newcommand{\Rbs}{\mbox{${{\scriptstyle \not}{\scriptscriptstyle R}}$}}

\draft

\title{Low Scale Leptogenesis and Dark Matter Candidates\\ in an Extended Seesaw Model}

\thispagestyle{empty}
\author{ H. Sung Cheon$^{1,}$\footnote{E-mail:
        hsungcheon@gmail.com},~~ Sin Kyu Kang$^{2,}$\footnote{E-mail:
        skkang@snut.ac.kr},~~ C. S. Kim$^{1,}$\footnote{E-mail:
        cskim@yonsei.ac.kr} }
\affiliation{$ ^{1}$ Department of Physics, Yonsei University, Seoul 120-749, Korea\\
             $ ^{2}$ School of Liberal Arts, Seoul National University of Technology,
                     Seoul 121-742, Korea }

\begin{abstract}
\noindent
We consider a variant of seesaw mechanism by introducing extra singlet neutrinos and singlet scalar boson,
and show how low scale leptogenesis is successfully realized in this scenario.
We examine if the newly introduced neutral particles, either singlet Majorana neutrino or singlet scalar
boson, can be a candidate for dark matter.
We also discuss the implications of the dark matter detection through the scattering off the nucleus
of the detecting material on our scenarios for dark matter.
In addition, we study the implications for the search of {\it invisible} Higgs decay at LHC,
which may serve as a probe of our scenario for dark matter.
\end{abstract}
 \maketitle \thispagestyle{empty}
%

\section{Introduction}

Two unsolved important issues in particle physics and cosmology are why there is more matter
than antimatter in the present Universe and what is the origin of dark matter.
In this paper, we propose a model to address both  of those problems and show that they can be solved by means of a common origin..

One of the most popular models to accommodate the right amount of baryon asymmetry in the present
Universe is so-called leptogenesis \cite{lepto}, which is realized in the context of seesaw mechanism
\cite{seesaw}, and thus
the smallness of neutrino masses and the baryon asymmetry can be simultaneously achieved.
However, the typical leptogenesis demands rather large scale of the right-handed Majorana neutrino mass,
which makes it impossible to probe in present experimental laboratories.
To lower the scale of leptogenesis, we have recently proposed a variant of seesaw mechanism
\cite{kk} and showed that the required leptogenesis is allowed at a low scale even a few TeV scale
without imposing the tiny mass
splitting between two heavy Majorana neutrinos required in the resonant leptogenesis \cite{reso}.
The main point of our model previously proposed is to introduce an equal number of  gauge singlet neutrinos in addition to the heavy right-handed singlet neutrinos. In our scenario, there exist new Yukawa interactions
mediated by singlet Higgs sector which is coupled with the extra singlet neutrinos and
the right-handed singlet neutrinos.
As shown in \cite{kk}, the new Yukawa interactions may play a crucial role in enhancing the lepton
asymmetry, so that low scale leptogenesis can be achieved.

On the other hand, it is worthwhile to examine if those new
particles, either the newly introduced singlet neutrinos or singlet
scalar bosons, can be a dark matter candidate because  any kind of
neutral, stable and weakly interacting massive particles (WIMPs) can
be regarded as a good candidate for dark matter. While light singlet
neutrinos with mass of order MeV or keV have been considered as a
warm dark matter candidate \cite{warm}, heavy singlet neutrinos with
mass of order 100 GeV as a dark matter candidate have not been
studied much in detail. While the cold dark matter (CDM) models
supplemented by a cosmological constant are in a good agrement with
the observed structure of the Universe on large scales, the cosmic
microwave background anisotropies and type Ia supernovae
observations for a given set of density parameters, there exists a
growing wealth of observational data which are in conflict with the
CDM scenarios \cite{cdm}. To remedy the difficulties of the CDM
models on galactic scales, a self-interacting dark matter candidate
has been proposed \cite{self, scalarDM, SDM1}, and it has been
discussed that a gauge singlet scalar coupled to the Higgs boson,
leading to an invisible decaying Higgs, is a good candidate for the
self-interacting dark matter \cite{sh, sh2, sh3}.

In this paper, we will show that either the new gauge singlet
neutrinos or singlet scalar bosons, which play an important role in
enhancing the lepton asymmetry so that the low scale leptogenesis is
realized, can be a good candidate for dark matter. For our purpose,
we will first present how the enhancement of the lepton asymmetry
through the mediation of the newly introduced particles in our
framework can be achieved and then will show that the new
gauge singlet particles with the parameter space in consistent with
low scale leptogenesis can be satisfied with the criteria on the
candidate for dark matter. We will investigate possibilities of dark
matter detection through the scattering off the nucleus of the
detecting material. In addition, we will study how we can probe our
scenarios at high energy colliders and present that the search for
the invisible Higgs decay may  serve as a probe of dark matter
properties.

\section{Extended Seesaw Model and Low Scale Leptogenesis}
We begin by explaining what the extended seesaw model is and examine
how low scale leptogenesis can be realized in this context. The
Lagrangian we propose is given in the charged lepton basis as \be
{\cal L}_f=Y_{\nu_{ij}} \bar{\nu}_i H  N_j+M_{R_{ij}}N_iN_i+
Y_{S_{ij}} \bar{N}_i\Phi S_j
  -m_{S_{ij}} S_i S_j +h.c. \label{lag1}
\ee
where $\nu_i,~N_i,~S_i$ stand for SU(2)$_L$ doublet, right-handed
singlet and newly introduced singlet neutrinos, respectively.
$Y_{\nu_{ij}}$, $Y_{S_{ij}}$,  $M_{R_{ij}}$ represent Dirac Yukawa coupling matrix,
Singlet Yukawa coupling matrix and Heavy Majorana neutrino mass matrix, respectively.
And $H$ and $\Phi$ denote the SU(2)$_L$ doublet and singlet Higgs
scalars. Here, we impose $Z_2$ symmetry under which  $S_i$ and
$\Phi$ are odd and all other particles even, which makes this model
different from the extended double seesaw model proposed in
\cite{kk} even though the contents of particles are the same.
The immediate consequence of the exact $Z_2$ symmetry is that
the lightest $S_i$ or $\Phi$ can be a candidate for the cold dark matter
of the Universe.

Due to the exact $Z_2$ symmetry, the singlet scalar
field $\Phi$ can not drive a vacuum expectation value. Thus, in this
model, light neutrino masses are generated by typical seesaw
mechanism, which makes this model different from the model in \cite{kk}.
After integrating out the right-handed heavy neutrino
sector $N_R$ in the above Lagrangian,
 the light  neutrino masses are given by
 \begin{eqnarray}
  m_\nu =\frac{(Y_{\nu}  v_{_{\rm EW}})^2}{4M_R}, \label{seesaw}
 \end{eqnarray}
where we omitted the indices of
the mass matrix and Dirac Yukawa coupling, and $v_{_{\rm EW}} = 246$ GeV
is the Higgs vacuum expectation value. Although the absolute values
of three neutrino masses are unknown, their
masses are expected to be of order of $\sqrt{\Delta m^2
_{atm}} \simeq 0.05$ eV and $\sqrt{\Delta m^2 _{atm}} \simeq 0.01$
eV, provided that the mass spectrum of neutrinos is hierarchical.
There is also a bound on neutrino masses coming from WMAP
observation, which is $m_\nu \lesssim 0.23$ eV.
Thus it is interesting to see how the neutrino
masses of order of $0.01 \sim 0.1$ eV can be obtained in our
scenario. Such light neutrino masses can be generated through seesaw formula Eq. (\ref{seesaw}),
if we take, as an
example, $Y_{\nu}$ and $M_R$ to be of order
$10^{-6}$ and $10^4$ GeV, respectively.

Now, let us consider how low scale leptogenesis can be achieved by
the decay of the lightest right-handed Majorana neutrino in our
framework.
Right handed heavy
Majorana neutrinos are even under $Z_2$ symmetry, so that they can decay into a
pair of the singlet particles, $\Phi$ and $S$, or the standard model Higgs
and the lepton doublet.
Since the Yukawa couplings $(Y_S)_{2(3)i}$ are taken to be large,
the processes involving $N_{2,3}$ remain in thermal equilibrium
even at $T\simeq M_{R_1}$, and thus the decays of $N_{2,3}$ can not
lead to the desired baryon asymmetry.
However, the decay processes of the lightest right-handed Majorana neutrino $N_1$ depart
from thermal equilibrium at $T\lesssim M_{R_1}$, and thus lead to the desired baryon asymmetry

It will be  shown that there exists a new contribution to
the lepton asymmetry which is mediated by the extra singlet
neutrinos $S_i$ and  scalar boson $\Phi$, and successful
leptogenesis can be realized with rather light right-handed Majorana
neutrino masses which can escape the gravitino problem encountered
in supersymmetric standard model. Without loss of generality, we can
rotate and rephase the fields to make the mass matrices $M_{R_{ij}}$
and $m_{S_{ij}}$ real and diagonal. In this basis, the elements of
$Y_\nu$ and $Y_S$ are in general complex. The lepton number asymmetry
required for baryogenesis is given by
\begin{eqnarray}\epsilon_{1} &=& -\sum_i\left[\frac{\Gamma(N_1
\to \bar{l_i}\bar{H}) - \Gamma(N_1 \to l_iH) }{\Gamma_{\rm
tot}(N_1)}\right] ,
\end{eqnarray}
where $N_1$ is the lightest right-handed neutrino and $\Gamma_{\rm
tot}(N_1)$ is the total decay rate. Thanks to the new Yukawa
interactions, there is a new contribution of the diagram which
corresponds to the self energy correction of the vertex arisen due
to the new Yukawa couplings with singlet neutrinos and Higgs
scalars.
\fig [t] \ct{\ep{figure=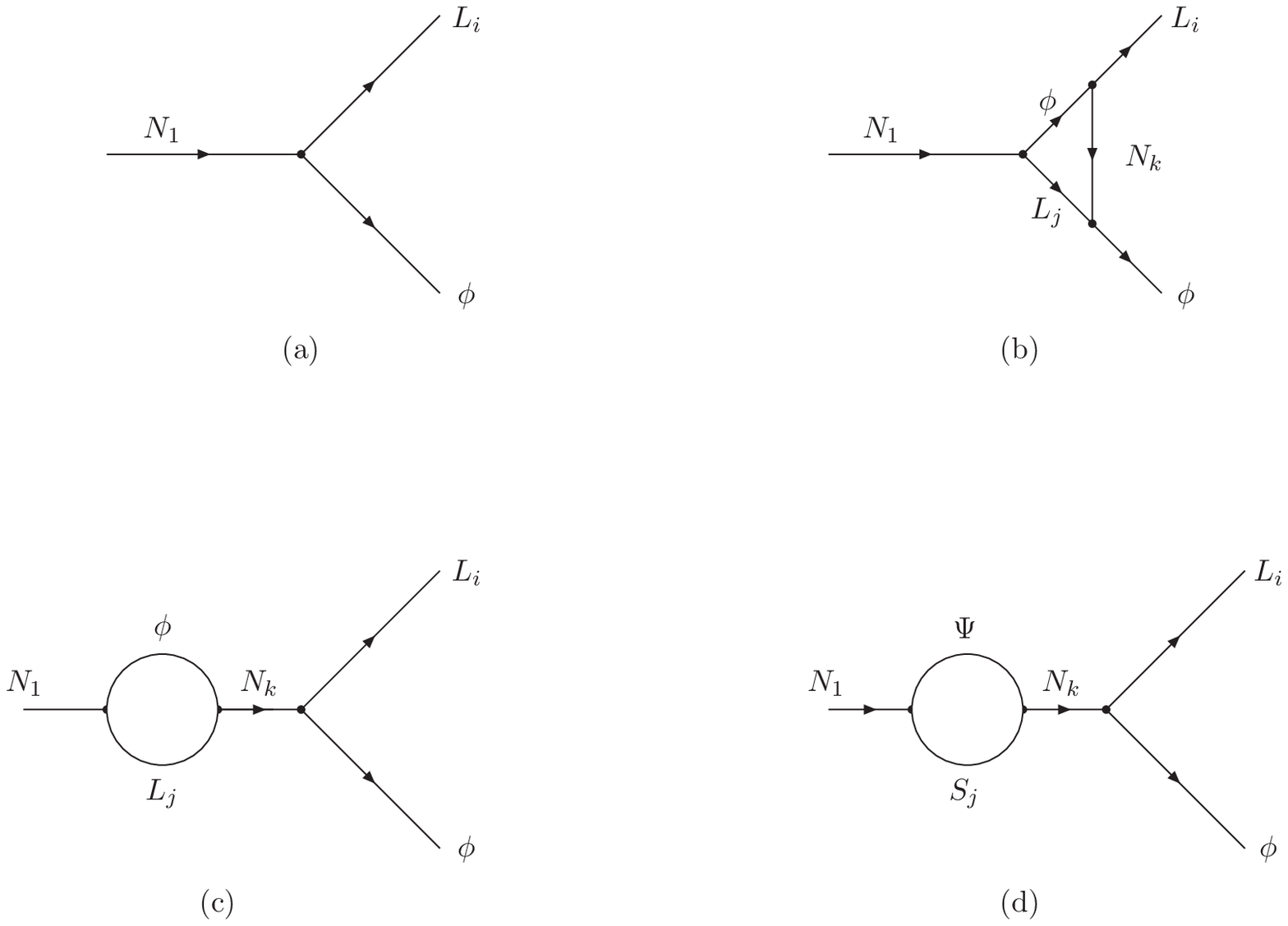, scale=0.5}} \cp{ Diagrams contributing to the lepton asymmetry.}\ef
Fig. 1 shows the structure of the diagrams contributing
to $\epsilon_1$. Assuming that the masses of the Higgs scalars and the newly
introduced singlet neutrinos are much smaller compared to that of
the right-handed neutrino, to leading order, we have
\begin{equation}
  \label{eq:vv}
\Gamma_{\rm tot}(N_i)={(Y_\nu Y_\nu^\dagger+Y_SY_S^\dagger)_{ii}
\over 4\pi}M_{R_i}
\end{equation}
so that
\begin{equation}
\epsilon_1 = \frac{1}{8\pi} \sum_{k\ne 1} \left( [ g_V(x_k)+
g_S(x_k)]{\cal T}_{k1} + g_S(x_k){\cal S}_{k1}\right),
\end{equation}
where $Y_{\nu}$ and $Y_S$ are given in the basis where $M_R$ and $M_S$ are diagonal,
$g_V(x)=\sqrt{x}\{1-(1+x) {\rm ln}[(1+x)/x]\}$,
$g_S(x)=\sqrt{x_k}/(1-x_k) $ with $x_k=M_{R_k}^2/M_{R_1}^2$ for
$k\ne 1$,
\begin{equation}
  \label{eq:vv}
{\cal T}_{k1}={{\rm Im}[(Y_\nu Y_\nu^\dagger)_{k1}^2]
 \over (Y_\nu Y_\nu^\dagger +Y_S Y_S^\dagger)_{11}}
\end{equation}
and
\begin{equation}
{\cal S}_{k1}={{\rm Im}[(Y_\nu Y_\nu^\dagger)_{k1}(Y_S^\dagger
Y_S)_{1k}]
 \over (Y_\nu Y_\nu^\dagger +Y_S Y_S^\dagger)_{11}}.
\end{equation}
Notice that the term proportional to ${\cal S}_{k1}$ comes from the
interference between the tree-level diagram with the new
contribution.

In particular, for $x\simeq 1$, the vertex contribution to
$\epsilon_1$ is much smaller than the contribution of the
self-energy diagrams and the asymmetry $\epsilon$ is resonantly
enhanced, and we do not consider this case.
 To see how much the new contribution may be important in this case, for simplicity,
 we consider a particular situation where $M_{R_1}\simeq M_{R_2} < M_{R_3}$,
 so that the effect of $N_3$ is negligibly small.
 In this case, the asymmetry 
 can be written as
 \begin{eqnarray}
 \epsilon_1 & \simeq & -\frac{1}{16\pi}\left[
           \frac{M_{R_2}}{v^2}\frac{Im[(Y^{\ast}_{\nu}m_{\nu}Y^{\dagger}_{\nu})_{11}]}
           {(Y_{\nu}Y^{\dagger}_{\nu}+Y_SY^{\dagger}_S)_{11}} \right. \nonumber \\
      & &     \left.+\frac{\sum_{k\ne 1} Im[(Y_{\nu}Y^{\dagger}_{\nu})_{k1}(Y_SY_S^{\dagger})_{1k}]}
                   {(Y_{\nu}Y^{\dagger}_{\nu}+Y_SY^{\dagger}_S)_{11}}\right]R~,
                   \label{epsilon2}
 \end{eqnarray}
where $R$ is a resonance factor defined by $R \equiv
|M_{R_1}|/(|M_{R_2}|-|M_{R_1}|)$. For successful leptogenesis, the
size of the denominator of $\epsilon_1$ should be constrained by the
out-of-equilibrium condition, $\Gamma_{N_1} < \mbox{H}|_{T=M_{R_1}}$ with
the Hubble expansion rate $\mbox{H}$, from which the corresponding upper
bound on the couplings $(Y_S)_{1i}$ reads
$\sqrt{\sum_i|(Y_S)_{1i}|^2}<3\times
10^{-4}\sqrt{M_{R_1}/10^9(\mbox{GeV})}$.
Then, the first term of Eq. (\ref{epsilon2}) is bounded as
$(M_{R_2}/16\pi v^2)\sqrt{\Delta m_{atm} ^2}R$. So if the first term of
Eq. (\ref{epsilon2}) dominates over the second one, $R\sim10^{6-7}$
is required to achieve TeV scale leptogenesis, which implies severe
fine-tuning. However, since the size of $(Y_S)_{2i}$ is not
constrained by the out-of-equilibrium condition, large value of
$(Y_S)_{2i}$ is allowed for which the second term of Eq.
(\ref{epsilon2}) can dominate over the first one and thus the size
of $\epsilon_1$ can be enhanced.
For example, if we assume that $(Y_\nu)_{2i}$ is aligned to $(Y_S ^*)_{2i}$, $i.e.$
 $(Y_S)_{2i} = \kappa (Y_\nu ^*)_{2i}$ with constant $\kappa$, the upper limit of the second term of
 Eq. (\ref{epsilon2}) is given in terms of $\kappa$ by $|\kappa|^2 M_{R_2}\sqrt{\Delta m^2 _{atm}}R/16\pi v^2$,
 and then we can achieve the successful low scale leptogenesis by taking rather large value of $\kappa$,
 instead of imposing very tiny mass splitting between $M_{R_1}$ and
 $M_{R_2}$.
The right amount of the asymmetry, $\epsilon_1\sim
10^{-6}$, can be obtained for $M_{R_1}\sim 10^4\, \mbox{GeV}$,
provided that $\kappa=(Y_S)_{2i}/(Y_{\nu})^{\ast}_{2i}\sim 10^{3}$
and $M^2 _{R_2}/M^2 _{R_1}\sim 10$.
We {\it emphasize} that
such a requirement for the hierarchy between $Y_{\nu}$ and $Y_S$ is
much less severe than the required fine-tuning of the mass splitting
between two heavy Majorana neutrinos to achieve the successful
leptogenesis at low scale.

The generated B-L asymmetry is given by  $Y^{\rm SM}_{\rm B-L}=-\eta
\epsilon_1 Y^{eq}_{N_1}$, where $Y^{eq}_{N_1}$ is the number density
of the right-handed heavy neutrino at $T \gg M_{R_1}$ in thermal
equilibrium given by $Y^{eq}_{N_1}\simeq
\frac{45}{\pi^4}\frac{\zeta(3)}{g_{\ast}k_B} \frac{3}{4}$ with
Boltzmann constant $k_B$ and the effective number of degree of
freedom $g^{\ast}$. The efficient factor $\eta$ can be computed
through a set of coupled Boltzmann equations which take into account
processes that create or washout the asymmetry. To a good
approximation the efficiency factor depends on the effective
neutrino mass $\tilde{m}_1$ given in the presence of the new Yukawa
interactions with the coupling $Y_S$ by
\begin{eqnarray}
\tilde{m}_1=\frac{(Y_{\nu}Y^{\dagger}_{\nu}+Y_SY^{\dagger}_S)_{11}
}{M_{R_1}}v^2.
\end{eqnarray}
In our model, the new process of type $S\Phi \rightarrow lH$ will
contribute to wash-out of the produced B-L asymmetry. The process
occurs mainly through virtual $N_{2,3}$ exchanges  because
the Yukawa couplings $(Y_S)_{2(3)i}$ are taken to be large in our
model and the rate is proportional to
$M_{R_1}|Y_SY_{\nu}^{\ast}/M_{R_{2,3}}|^2$.
Effect of the wash-out can be easily estimated from the fact that it
looks similar to the case of the typical seesaw model if  $M_{R_1}$
is replaced with $M_{R_1}(Y_S/Y_{\nu})^2$. It turns out that the
wash-out factor for $(Y_S)_{1i}\sim (Y_\nu)_{1i}$,
$(Y_S)_{2i}/(Y_{\nu})_{2i}\sim 10^3$ and $M_{R_1}\sim 10^4$ GeV is
similar to the case of the typical seesaw model with  $M_{R_1}\sim
10^4$ GeV and $\tilde{m}_1\simeq 10^{-3}$ eV, and is estimated so
that $\epsilon_{1} \sim 10^{-6}$ can be enough to explain the baryon
asymmetry of the Universe provided that the initial lightest right-handed neutrino is thermal \cite{Buchmuller}.


\section{Investigation of possible Dark Matter Candidates}

Now, let us examine if either the newly introduced singlet Majorana neutrino or
the singlet scalar boson can be a candidate for dark matter.
For our purpose, in addition to the lagrangian ${\cal L}_f$ given in Eq. (\ref{lag1}),
we allow quartic scalar interactions for the scalar sectors.
Then, the lagrangian we consider is given by
\be
 {\cal L}&=&{\cal L}_f +\frac{1}{2}\left(\partial_{\mu} \Phi \right)^2- \frac{1}{2}m^2_{\Phi^0} \Phi ^2
 - \frac{\lambda_s}{4} \Phi^4
- \lambda H^{\dagger}H \Phi^2. ~\label{lg2}
\ee
Note that the self-interacting coupling of the singlet scalar $\lambda_s$ is
largely unconstrained and thus can be chosen arbitrarily. But, we assume
that it should not be so large that the perturbation breaks down.
As mentioned before, in order to guarantee the stability of the dark matter candidate, we
impose $Z_2$ discrete symmetry under which all the standard model (SM) particles are even whereas the singlet
Majorana neutrinos $S_i$ and singlet scalar boson $\Phi$ are odd.
In addition, we demand that the scalar potential
is bounded from below so as to guarantee the existence of a vacuum
and the minimum of the scalar potential must spontaneously break the
electroweak gauge group, $<H^0>\neq 0$, but must not break $Z_2$
symmetry imposed above.
After breaking the electroweak gauge symmetry, the singlet scalar $\Phi$- dependent part
of the scalar potential is given by
\begin{eqnarray}
V=\frac{1}{2}(m^2_{\Phi^0}+\lambda v_{_{\rm EW}}^2)\Phi^2
+\frac{\lambda_s}{4}\Phi^4+ \lambda v_{_{\rm EW}} \Phi^2 h+
\frac{\lambda}{2}\Phi^2 h^2, ~\label{sp}
\end{eqnarray}
where we have adopted $\sqrt{2}H^{\dagger}=(h,0)$ and shifted the
Higgs boson $h$ by $h\rightarrow h+v_{_{\rm EW}}$, where $v_{_{\rm EW}}=246$
GeV is the Higgs vacuum expectation value. The physical mass of
$\Phi$, $m_{\Phi}$, is then given by $m^2_{\Phi^0}+\lambda
v_{_{\rm EW}}^2$. We assume that the spectrum of the singlet neutrinos
$S_i$ is not degenerate and  two heavier ones $S_{2,3}$ are so much
heavier than {\bf $S_1$}  that  they could not be dark matter
candidates. Here, we notice that only the lightest odd particle
(LOP) under $Z_2$ can be a candidate for dark  matter in our
scenario because the next lightest odd particle (NOP) can be decayed
into the LOP by the cascade decay, as shown in Fig. 2.

\fig [t]
\ct{\ep{figure=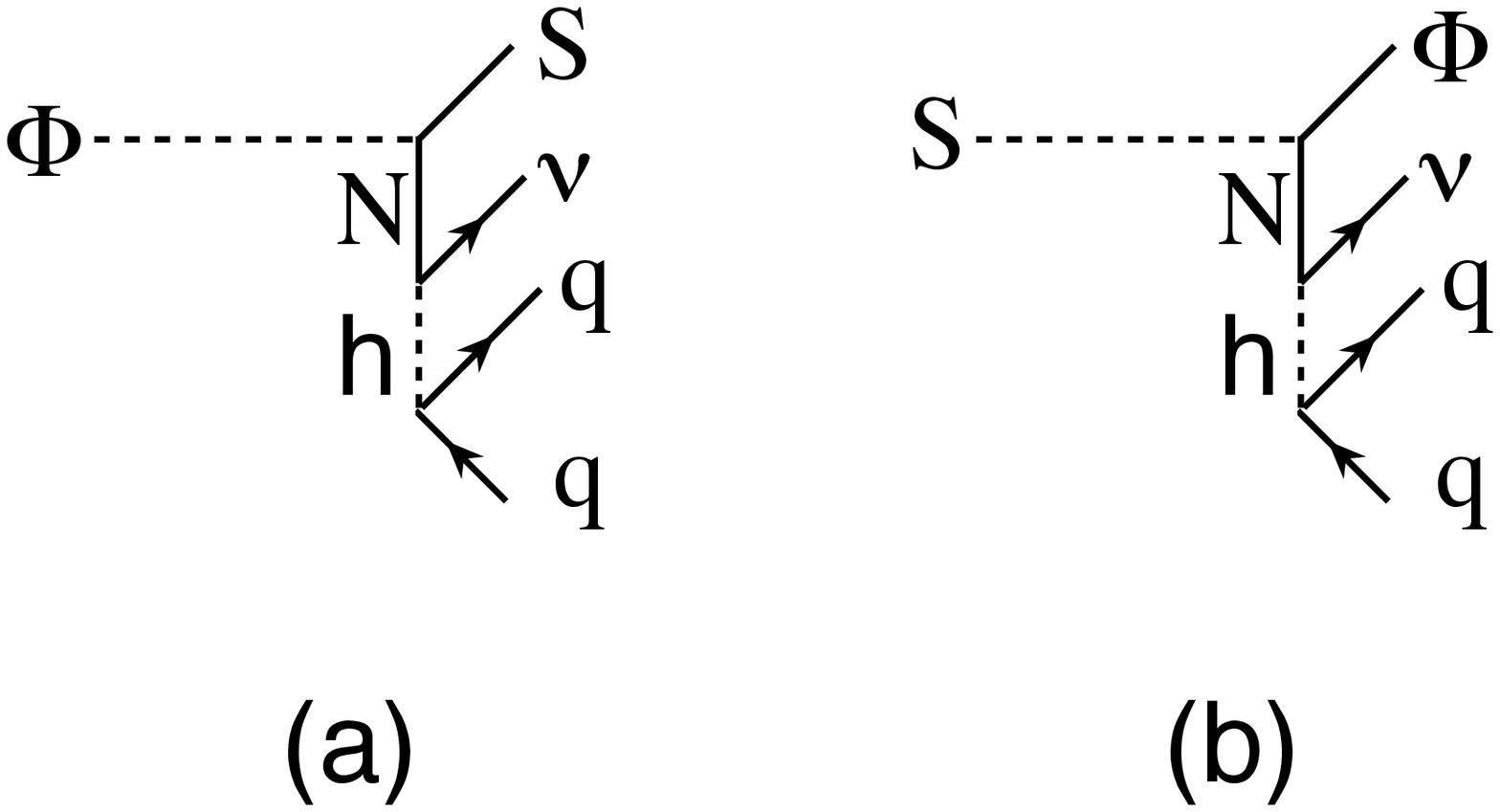, scale=0.4}} \cp{ Cascade decays of NOP to
LOP : (a) case for $m_S < m_{\Phi}$, (b) case for $m_{\Phi}< m_S$
}\ef

Since there are two kinds of odd particles under $Z_2$ in our scenario,
we can classify two possible candidates for dark matter according to which particle is the LOP.
One is the case that the singlet Majorana neutrinos $S_1$ is the LOP and the other is that $\Phi$ is the LOP.
As will be clearer later, the primordial abundance of dark matter candidate is predicted as a function of
masses and coupling $\lambda$.
Thus, imposing the preferred values of $\Omega_{DM} h^2$ observed from WMAP,
$0.094<\Omega_{DM} h^2 <0.128$ \cite{CMB1,CMB2}, we can get a strong relation between mass of dark matter candidate
and coupling $\lambda$.

\subsection{ Singlet Neutrino $S$ as a Dark Matter Candidate}

The singlet neutrino $S_1$  can be a dark matter candidate, provided
that $m_{S_1} \lesssim m_{\Phi}$. We omit the generation index $1$
of $S_1$ hereafter. To estimate the relic abundance of the singlet
neutrino $S$ at a freeze-out temperature $x_f=m_S/T_f$ , we need to know the annihilation processes\cite{EU}. There are two possible
annihilation processes of the singlet neutrino, one is happened at
loop level and the other is mediated by $\Phi$.
But, both possibilities are not relevant
to fit the required present relic density of the dark matter because
the annihilation cross section for the dark matter candidate $S$ is too small, therefore,
these processes would predict too much relic abundance of $S$.
However, it is well known that if
the mass of NOP is close to that of LOP, it would not be decoupled
from thermal equilibrium at the freeze-out temperature of the LOP
and thus influences the relic abundance of LOP \cite{coan}. We call
this {\it coannihilation} \cite{coan} and this mechanism lowers the present day
relic density of $S$ in this scenario.
It turns out that  in our scenario if $\delta m = m_{\Phi} - m_S \approx T_f$,
annihilation processes of the NOP into
a pair of the SM particles through the s-channel,
as shown in Fig. 3, can significantly affect the relic abundance of
$S$ to be appropriately reduced.

\fig [t] \ct{\ep{figure=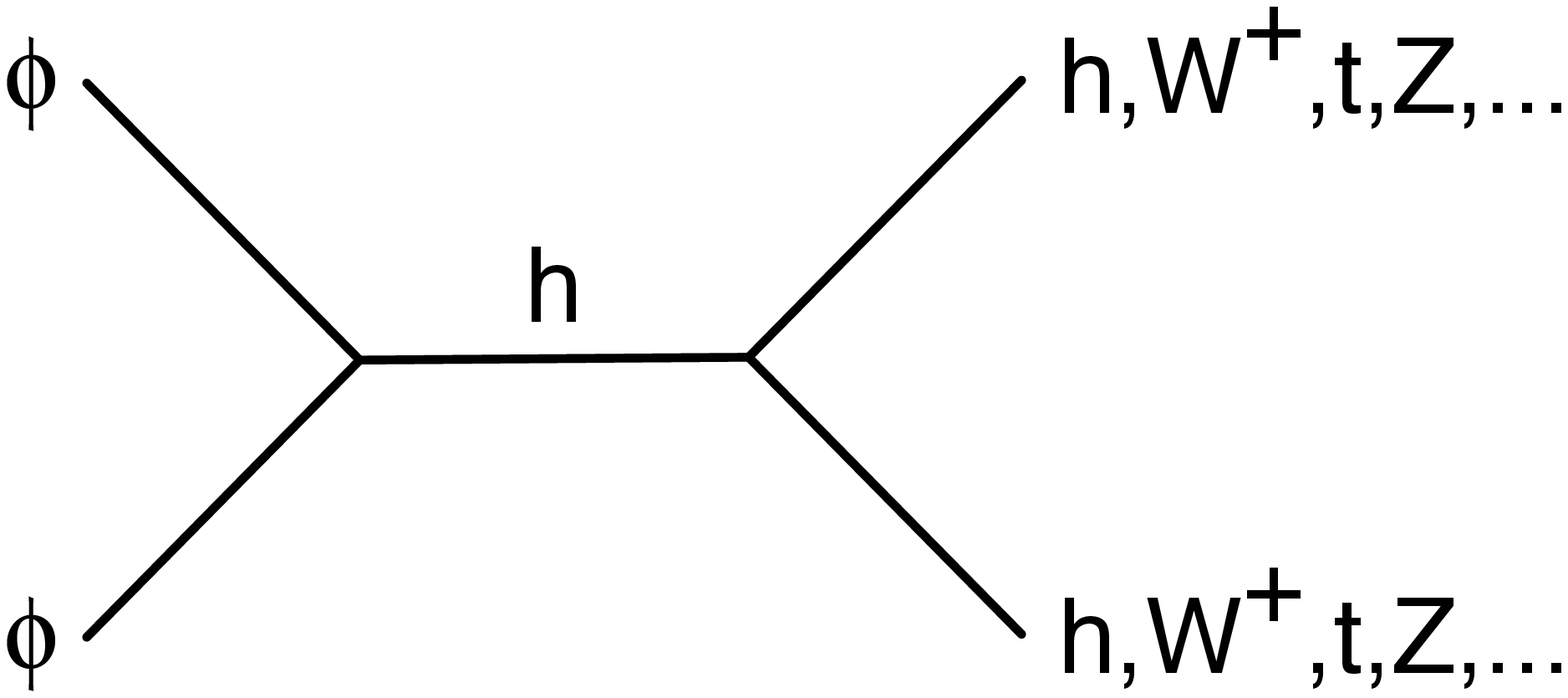, scale=0.4}}\cp{Diagram for
annihilation processes of the singlet scalar bosons $\Phi$} \ef

With the help of the standard formulae for calculating relic abundance of $s$-wave annihilation
\cite{EU},
\be
\Omega_S h^2 = \frac{(1.07\times 10^9) x_f \mbox{GeV}^{-1}}{g^{1/2} _* M_{pl}\int^{\infty}_{x_f}
                 <\sigma_{{\rm eff}}v_{\rm rel}>x^{-2}dx}, \label{omega}
\ee
we can estimate the present relic density of singlet Majorana neutrino $S$.
Here $g_*$ is the degrees of freedom in equilibrium at annihilation, and $x_f=m_{\rm LOP}/T_f$ is the inverse freeze-out
temperature in units of $m_{\rm LOP}$,  which can be determined by solving the equation
\begin{eqnarray}
x_f\simeq \ln \left(0.038g_{{\rm eff}}(g_{\ast}\cdot x_f)^{-1/2}M_{Pl}m_{\rm LOP}<\sigma_{{\rm eff}} v_{{\rm rel}}>\right), \label{xf}
\end{eqnarray}
where $< >$ means the relevant thermal average,
$g_{{\rm eff}}=\sum_i g_i(m_i/m_{\rm LOP})^{3/2}e^{-(m_i-m_{\rm LOP})/T} $ with the number of degree of freedom $g_i$
for $i=(S, \Phi)$,
and $\sigma_{{\rm eff}}$ is the effective cross section defined in \cite{coan}.
On calculating $\Omega_S h^2$, we have used the \emph{micrOMEGAs 2.0} program \cite{micro}.
The dominant contribution to $\sigma_{{\rm eff}}$ in this case is the pair annihilation cross section of
the heavier particle $\Phi$.
Since it is $T_f \sim 0.04\, m_\Phi$ that the annihilation of $\Phi$ is important, what we need to know is
the non-relativistic annihilation cross section.  In the non-relativistic limit, s-channel annihilation of $\Phi$
via Higgs exchange is given by \cite{sh},
\begin{eqnarray}
\sigma_{ann}v_{\rm rel}=\frac{8\lambda^2v^2_{_{\rm EW}}}{(4m^2_{\Phi}-m^2_h)^2+m^2_h\Gamma_h^2}F_X,
\end{eqnarray}
where $\Gamma_h$ is the total Higgs decay rate, and $F_X=\lim_{m_{\tilde{h}}\rightarrow 2m_{\Phi}}
\left(\frac{\Gamma_{\tilde{h}X}}{m_{\tilde{h}}} \right)$ with the partial rate for decay $\tilde{h}\rightarrow X$, for a virtual Higgs $\tilde{h}$.
Requiring $\Omega_S h^2$ to be in the region measured from WMAP,
$0.094<\Omega h^2 <0.128$ \cite{CMB1, CMB2}, we can obtain a relation between the coupling $\lambda$ and
the mass of the scalar boson $m_{\Phi}$.

\fig [b] \ct{\ep{figure=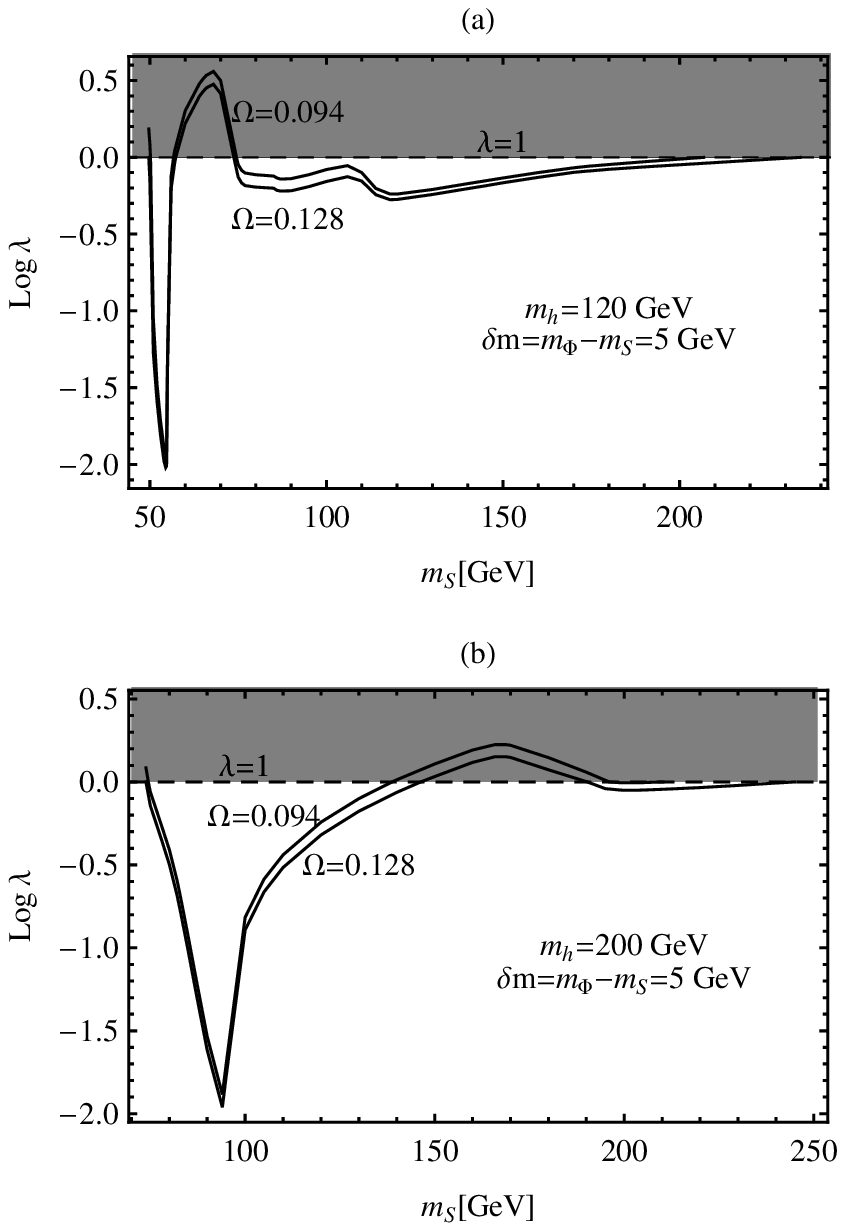, scale=0.7}} \cp{ Relationship
between $\lambda$ and $m_S$ arisen from the constraints $\Omega_S
h^2 = 0.128$ and $0.094$ corresponding to the upper and lower limit
of $\Omega_{DM}h^2$ measured from WMAP, respectively. Here the mass
difference $\delta m = m_{\Phi} -m_S$ has been taken to be 5 GeV and
Higgs mass $m_h$ to be (a) $120\, \mbox{GeV}$ and (b) $200\,
\mbox{GeV}$. Here the shadowed region is forbidden due to the breakdown
of perturbation.}\ef

In the case of $\Omega_S h^2 = 0.128$, in Fig. 4 we represent the relation between $\lambda$ and $m_S$ for
$ \delta m = 5\, \mbox{GeV}$.
Fig. 4-(a) corresponds to  the Higgs mass $m_h=120$ GeV, whereas
Fig. 4-(b) corresponds to  $m_h=200$ GeV.
The shadowed region is forbidden due to the breakdown of perturbation, so
the region $57~(146)~ \mbox{GeV} \leq m_{S}\leq 74 ~(190)$ GeV for  $m_h=120 ~(200)$ GeV and $\Omega h^2 =0.128$ is not relevant to our scenario.
In the case of $\Omega h^2 =0.128$, it is
$57 ~(139)~ \mbox{GeV} \leq m_{S}\leq 75 ~(196)$ GeV for  $m_h=120 ~(200)$ GeV.

We notice from Fig. 4 that
there exist kinematically special regions, such as the Higgs threshold ($2m_\phi \simeq m_H$)
and two-particle threshold in the final states ($m_\phi \simeq m_Z$ or $m_\phi \simeq m_H$, and so on).
We see from Fig. 4 that for
$50 ~(74)~ \mbox{GeV}\lesssim m_S \lesssim 234 ~(245)$ GeV for $m_h=120~(200)$ GeV and $\Omega h^2=0.128$ except the regions corresponding
to the poles and particle thresholds,
the abundance constraint arisen from WMAP results requires $\lambda \sim \mc{O}(0.5 (0.3) -1)$. This result indicates that
we do not need any fine tuning or special choice of the parameters in order to achieve right amount of relic abundance
of dark matter candidate.
It also turns out that if $m_S$ is lighter than $50 ~(74) $ GeV or heavier than $234 ~(245)$ GeV for
$m_h=120 ~(200)$ GeV and $\Omega h^2 =0.128$,
the relic abundance of $S$ is incompatible with WMAP results for the relic density.
It is worthwhile to notice that the coupling $\lambda$ gets significantly suppressed down to the level of $10^{-2}$ near the Higgs pole.
This is because the Higgs resonance is quite narrow, which in turn considerably enhances the scalar boson $\Phi$ annihilation rate,
especially if $2m_\Phi$ is slightly smaller than $m_h$ \cite{coan, sh}

\subsection{ Singlet Scalar Boson $\Phi $ as a Dark Matter Candidate:}

The singlet Higgs scalar $\Phi$ can be a candidate for dark matter, provided that $m_{\Phi}\lesssim m_S$.
Since $\Phi$ is the lightest particle involved in the interaction term $Y_S N S \Phi$,
this particle can not decay into other particles, and thus the annihilation processes relevant to a
successful candidate for dark matter can occur through the Higgs interaction term $\lambda \Phi^2 H^\dagger H$.
In this case, the singlet scalar bosons annihilate into the SM particles mediated by
the Higgs boson $h$.

In Ref. \cite{scalarDM}, it has been proposed that a stable,
strongly self-coupled scalar field can solve the problems of
cold dark matter models for structure formation in the Universe,
concerning galactic scales. Thus, the singlet scalar boson in our
model can serve as a self-interacting scalar dark matter candidate.
If the mass difference between $m_{\Phi}$ and $m_S$ is  substantially
large, the new Yukawa interactions $Y_S N S \Phi$ will negligibly affect
the relic density of $\Phi$, so the behavior of $\Phi$ as a dark
matter candidate is the same as that of the self-interacting scalar
dark matter candidate.  If the mass difference  between $S$ and
$\Phi$ is substantially small, the particles $S_i$ are thermally
accessible and they are as abundant as $\Phi$.
In this case, the formulae for the relic abundance of the particle $\Phi$ and its freeze-out
temperature $x_f$ are given by the same forms of  Eq. (\ref{omega}) and Eq. (\ref{xf}), respectively.
In fact, however, the annihilation cross sections associated with the heavier particles
$S_i$  are turned out to be negligibly small in this scenario, so $\sigma_{{\rm eff}}\simeq \sigma_{\Phi}
g^2/g^2_{{\rm eff}}$, where $\sigma_{\Phi}$ is the pair annihilation cross section
of $\Phi$ and $g$ is the internal degree of freedom of $\Phi$.
Inserting $\sigma_{{\rm eff}}$ into $x_f$, we see that the value of $x_f$ for $\delta m=m_S-m_{\Phi}=5$ GeV
 and $m_{\Phi}=500$ GeV in this scenario
is about $4\%$ lower than that in the singlet Higgs model without the particles $S_i$,
which in turn increases the relic abundance of $\Phi$.

\fig [t]
\ct{\ep{figure=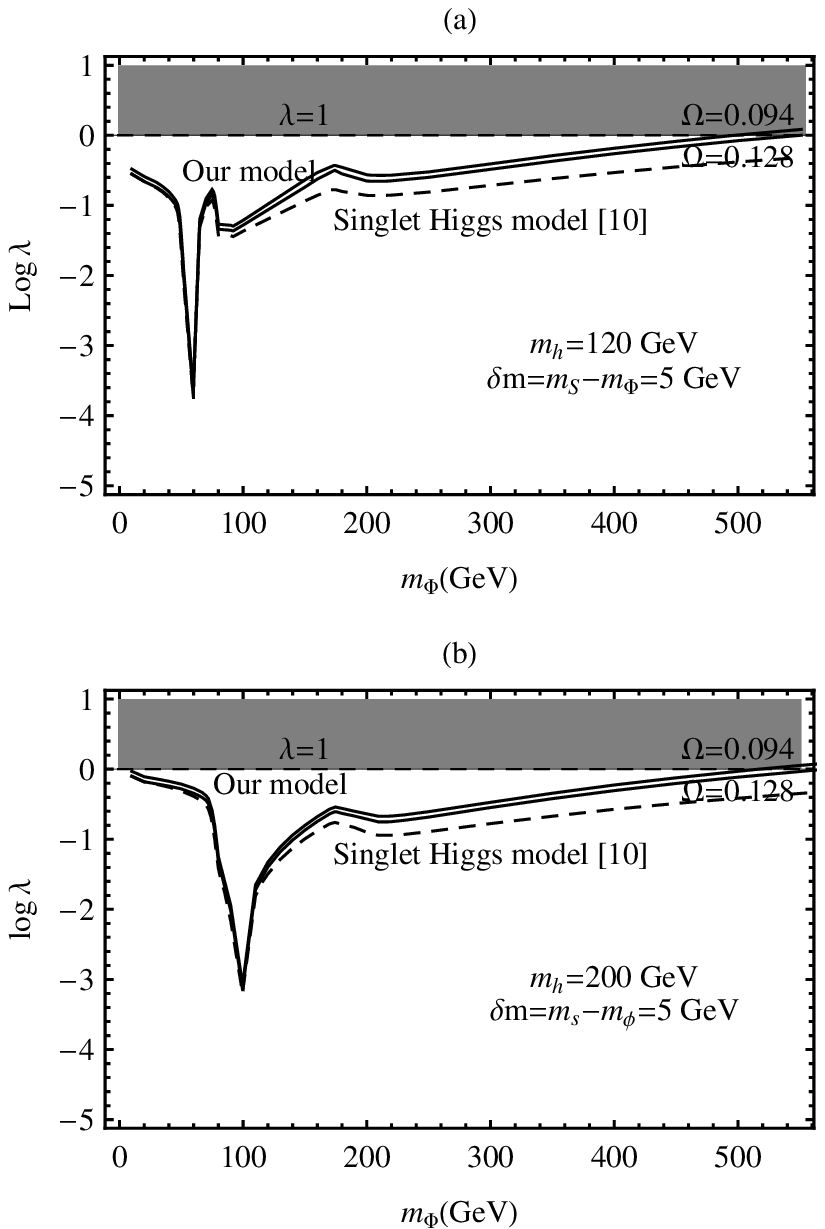, scale=0.7}} \cp{Relationship between $\lambda$ and $m_{\Phi}$
corresponding to $\Omega_\Phi h^2 =0.128$ (the lower solid line) and 0.094 (the upper solid line), respectively.
The mass difference $\delta m = m_S -m_\Phi$ has been taken to be 5 GeV and $m_h$ to be (a) 120 GeV
and (b) 200 GeV. Here the shadowed region is forbidden due to breakdown of perturbation and the dashed line
corresponds to the prediction of the model with a self-interacting dark matter \cite{sh}
for $\Omega_\Phi h^2 = 0.128$.}
\ef

Since several important features of our results are quite similar to those of \cite{sh}, therefore,
here we just present what are the differences between \cite{sh} and our model.
In Fig. 5, we show the relation between $\lambda$ and $m_{\Phi}$ for $m_h=120 ~(200)$ GeV and
$\delta m = m_S -m_\Phi =5$ GeV,
which is generated by requiring $\Omega_\phi h^2 =0.128$ and 0.094.
We see from Fig. 5 that  $m_{\Phi}$ should be less than 551 GeV ~(571 GeV)
for $m_h=120~(200)$ GeV and $\Omega h^2 =0.128$ in our model,
so as to be consistent with the relic abundance constraint
without breaking down perturbation ({\it i.e.} $\lambda \lesssim 1$).
This upper limit on $m_{\Phi}$ is much more restrictive than what is obtained in \cite{sh}.
Also, we see in Fig. 5-(a) that
when $m_{\Phi}$ lies between 80 GeV and 551 GeV, the value of $\lambda$ in our model  is much
larger than that given in \cite{sh}.

\section{Implication for Dark Matter search}

To directly detect dark matter, typically proposed method is  to detect the scattering of dark matter
off the nucleus of the detecting material. Since the scattering cross section is expected to be very small,
the energy deposited by a candidate for dark matter on the detector nucleus is also very small.
In order to measure this small recoil energy, typically of order keV, of the nucleus, a very low threshold detector
condition is required.
Since the sensitivity of detectors to a dark matter candidate is controlled by their elastic scattering cross section
with nucleus, it is instructive to examine how large the size of the elastic cross section could be.
First, to estimate the elastic cross section with nucleus, we need to know the relevant matrix element for
slowly moving spin-J nuclei, which is approximately given by \cite{sh}
\begin{eqnarray}
\frac{1}{2J+1}\sum_{spins}|<n^{\prime}|\sum_{f} y_f \bar{f}f|n>|^2 \simeq \frac{|A_n|^2}{(2\pi)^6},
\end{eqnarray}
where $n$ denotes nucleons and $|A_n|$ is determined to be
\be
\mc{A}_{n} = g_{_{Hnn}} \simeq \frac{190\, {\rm MeV}}{v_{_{\rm EW}}}  ~\label{an}
\ee
 by following the method given in \cite{sh} and taking the strange quark mass to be 95 MeV and
 $<n|\bar{s}s|n>\sim 0.7$.

Now, let us estimate the sizes of the elastic scattering cross sections in each case of dark matter candidates.
\\

{\bf  (i) ~ Case for $m_S < m_\phi$ :}\\
The Feynman diagram describing the scattering of the singlet Majorana neutrino $S$ with nucleons and nuclei is given
by $t$-channel Higgs and heavy Majorana neutrino exchange, as shown in Fig. 6-(a).
In this case, the non-relativistic spin-independent quasi-elastic scattering cross section is approximately given by
\be \sigma_{el} \approx \frac{Y^2 Y_\nu ^2 |A_n|^2}{128 \pi^3}
\Big ( \frac{m^2 _\nu m^2 _n}{m^4 _h m^2 _N m_S} \Big )\times m_{PL}, \ee
where $m_{PL}$ is the Planck mass.

\fig [t] \ct{\ep{figure=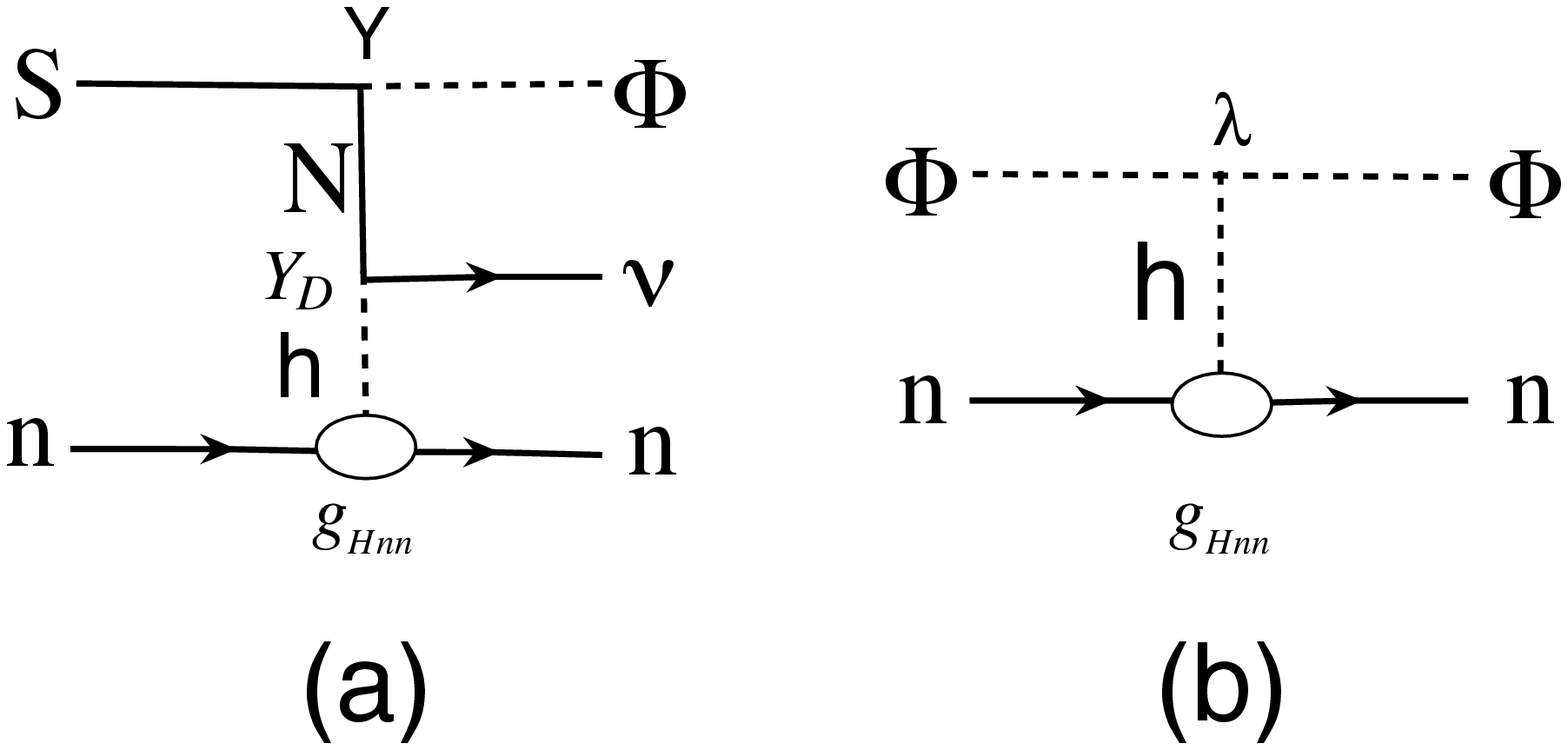, scale=0.4}}\cp{ Feynman diagrams relevant to (a) $S$-nucleon elastic scattering,
   and (b) $\Phi$-nucleon elastic scattering. }\ef

However, the size of $\sigma_{el}$ is turned out to be less than $10^{-71}\, cm^2$ due to small neutrino mass $m_{\nu}$
as well as small Yukawa couplings $Y$ and $Y_\nu$, which is much smaller than the current bound from
dark matter experiments. Thus, it is extremely difficult to detect the signal for the singlet Majorana neutrino $S$
at dark matter detectors. \\

{\bf (ii)~ Case for $m_\phi < m_S$:}\\
In this case, the Feynman diagram relevant to scalar-nucleon elastic scattering is presented in Fig. 6-(b), which
has already been considered in \cite{sh}.
Then, the non-relativistic elastic scattering cross section is given by \cite{sh}
\be \sigma_{el} = \frac{\lambda^2 v^2_{_{\rm EW}} |\mc{A}_n|^2}{4\pi} \Big (\frac{m^2_*}{m^2 _\phi  m^4 _h} \Big ),
\label{elasX}
\ee
where $m_* = m_\phi m_n / (m_\phi +m_n)$ is the reduced mass for the collision. Substituting Eq. (\ref{an}) into Eq. (\ref{elasX}),
\be \sigma_{el}(nucleon) &\approx& \frac{1}{4\pi}\Big (\frac{\lambda 190 \, MeV}{m^2 _h} \Big )^2 \Big ( \frac{m_p}{(m_p + m_\phi)} \Big )^2 \nn \\
&=& \lambda^2 \Big (\frac{100\, \mbox{GeV}}{m_h} \Big )^4 \Big ( \frac{50\, \mbox{GeV}}{m_\phi} \Big )^2 (4.47 \times 10^{-42}\, cm^2 ).
\nn 
\ee
where the mass of $m_p$ is a mass of proton.
In Fig. 7, we plot the predictions for the elastic scattering cross section as a function of the scalar mass $m_{\Phi}$ for
 $m_h=120\, \mbox{GeV}$ and $m_h=200\, \mbox{GeV}$, respectively, and the mass difference ($\delta m = m_S - m_\phi$) is taken to be $5\, \mbox{GeV}$.
The lower line corresponds to $\Omega_\Phi h^2 = 0.128$, whereas the upper line to $\Omega_\Phi h^2 = 0.094$.
On calculating $\sigma_{el}$, we used the relationship between $\lambda$ and $m_{\Phi}$ which is obtained through
the constraint from WMAP result as before.

\fig [t]
\ct{\ep{figure=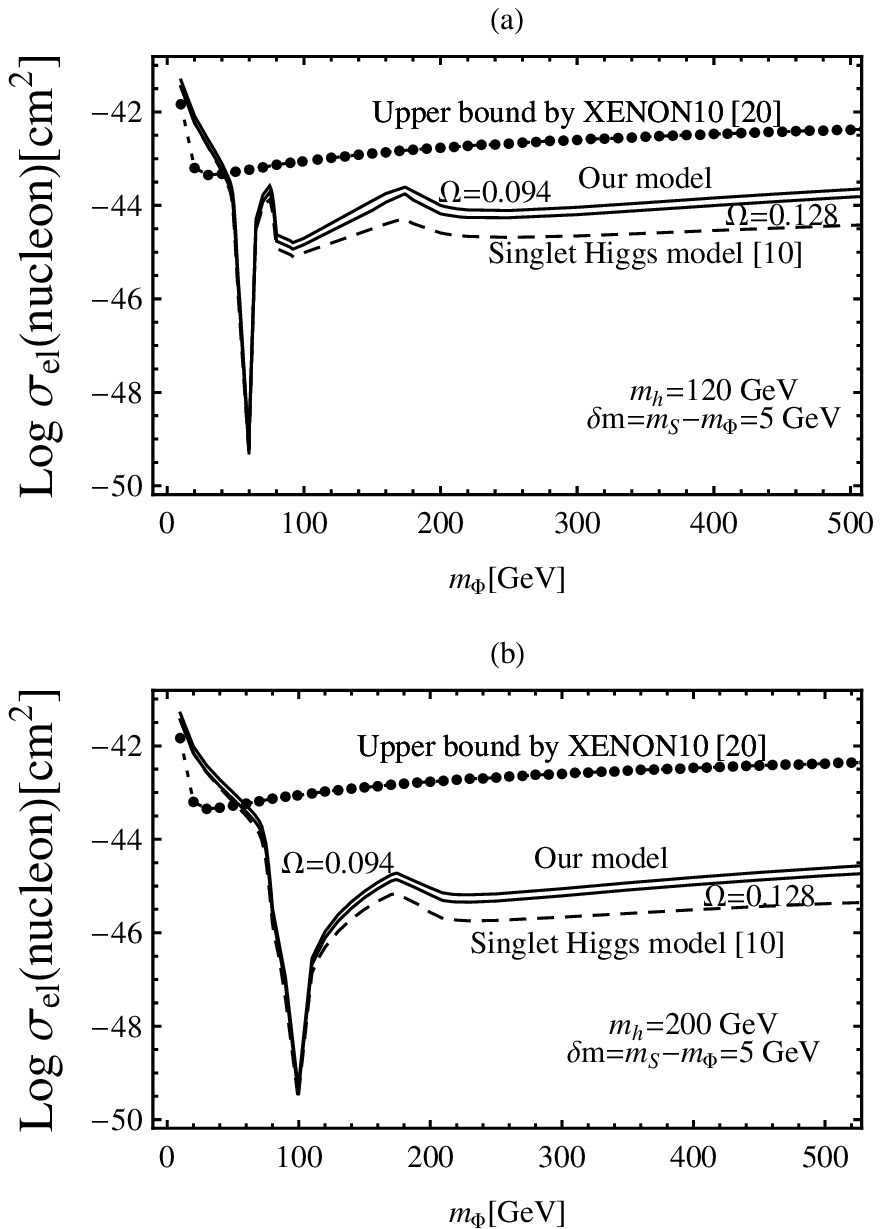, scale=0.7}}\cp{Plots of the elastic cross section $\sigma_{el}$
as a function of $m_{\Phi}$ for (a) $m_h=120$ GeV and (b) for $m_h=200$ GeV.
The dotted line is spin-independent WIMP-nucleon cross section upper limits (90\% C.L.)
by XENON10 Dark Matter Experiment \cite{xenon}. Here the dashed line corresponds
to the prediction of the model with a self-interacting dark matter \cite{sh} for $\Omega h^2 =0.128$.}
\ef

 In Fig. 7, we plot the new $90\, \%$ C.L. upper bound for the WIMP-nucleon spin-independent cross section
 as a function of $m_{\Phi}$ obtained from XENON10 Dark Matter Experiment \cite{xenon}.
As one can see from Fig. 7, when $\Omega h^2 =0.128$, the region $m_{\Phi}<42 ~(56)$ GeV for $m_h=120 ~(200)$ GeV and $\delta m=5$ GeV is
excluded  by XENON10 Dark Matter Experiment \cite{xenon}.

Comparing our results with those in the model of self-interacting scalar dark matter \cite{sh}, we see that
our prediction for the elastic cross section in the region $80 ~(110)~ \mbox{GeV} < m_{\Phi} < 551 ~(571)~ \mbox{GeV}$
for $m_h=120~ (200)$ GeV, and $\delta m=5$ GeV is about 1.4 (1.5) to 4.3 (4.4) times larger than that estimated in \cite{sh}.
This indicates that our scenario for scalar dark matter is distinguishable from the original model of the self-interacting scalar dark matter.

\section{ Implication for  Higgs searches at LHC}

Now we investigate the implications of our scenarios for Higgs searches at collider experiments.
The singlet scalar boson will not directly couple to ordinary matters, but only to the Higgs fields.
Although the presence of the singlet scalars will not affect electroweak phenomenology in a significant way,
it will affect the phenomenology of the Higgs boson.
Due to the large values for the coupling $\lambda \sim {\cal O}(0.1-1.0) $ required by relic abundance constraints,
real or virtual Higgs production may be associated with the singlet
Higgs $\Phi$ production, as discussed in \cite{sh}.
We see from Eq. (\ref{sp}) that if $2m_{\Phi}<m_h$, the real Higgs boson can decay into a pair of singlet scalars, whereas
if $2m_{\Phi}>m_h$, the singlet scalar bosons can not be produced by real Higgs decays, but arise only via virtual Higgs exchange.
As we know that any produced singlet scalar bosons are not expected to interact inside the collider,
thus they only give rise to strong missing energy signals.\\

{\bf  (i) Case for $2m_{\Phi} < m_h$ }: \\
In this case, the Higgs boson can invisibly decay into a pair of the singlet scalar bosons.
The invisible decay width is given at tree level by
\be \Ga_{H\rightarrow \phi \phi} = \frac{\lambda^2 v^2 _{_{\rm EW}}}{32\pi m_h} \sqrt{1-\frac{4m^2 _\Phi}{m^2_h}}. \ee
Since the relic abundance constraints require the large value of the coupling $\lambda$,
the invisible decay width to the Higgs boson gets large.
It is known that if the Higgs mass is less than $2M_W$ so that the Higgs partial width into the
SM particles is very small, the Higgs will decay predominantly into the singlet scalar bosons.
Then, LHC may yield a discovery signal for an invisible Higgs with enough reachable luminosity, for instance, 10 $fb^{-1}$ of
integrated luminosity for $m_h=120$ GeV, in associated production with Z boson \cite{xx}.
To quantify the signals for the invisible decay of the Higgs boson, we investigate the ratio $R$
defined as follows \cite{sh}:
\be R=\frac{Br_{h\rightarrow \bar{b}b,\, \bar{c}c, \, \bar{\tau}\tau} ({\rm SM}+\Phi)}{Br_{h\rightarrow \bar{b}b,\,
 \bar{c}c, \, \bar{\tau}\tau} ({\rm SM})} =
\frac{\Ga_{h,total} ({\rm SM})}{\Ga_{h\rightarrow \phi\phi} + \Ga_{h, total} ({\rm SM})}. \nn
\ee
The ratio $R$ indicates how the expected signal for the visible decay of the Higgs boson can decrease due to the existence of the singlet scalar bosons.

\fig [t]
\ct{\ep{figure=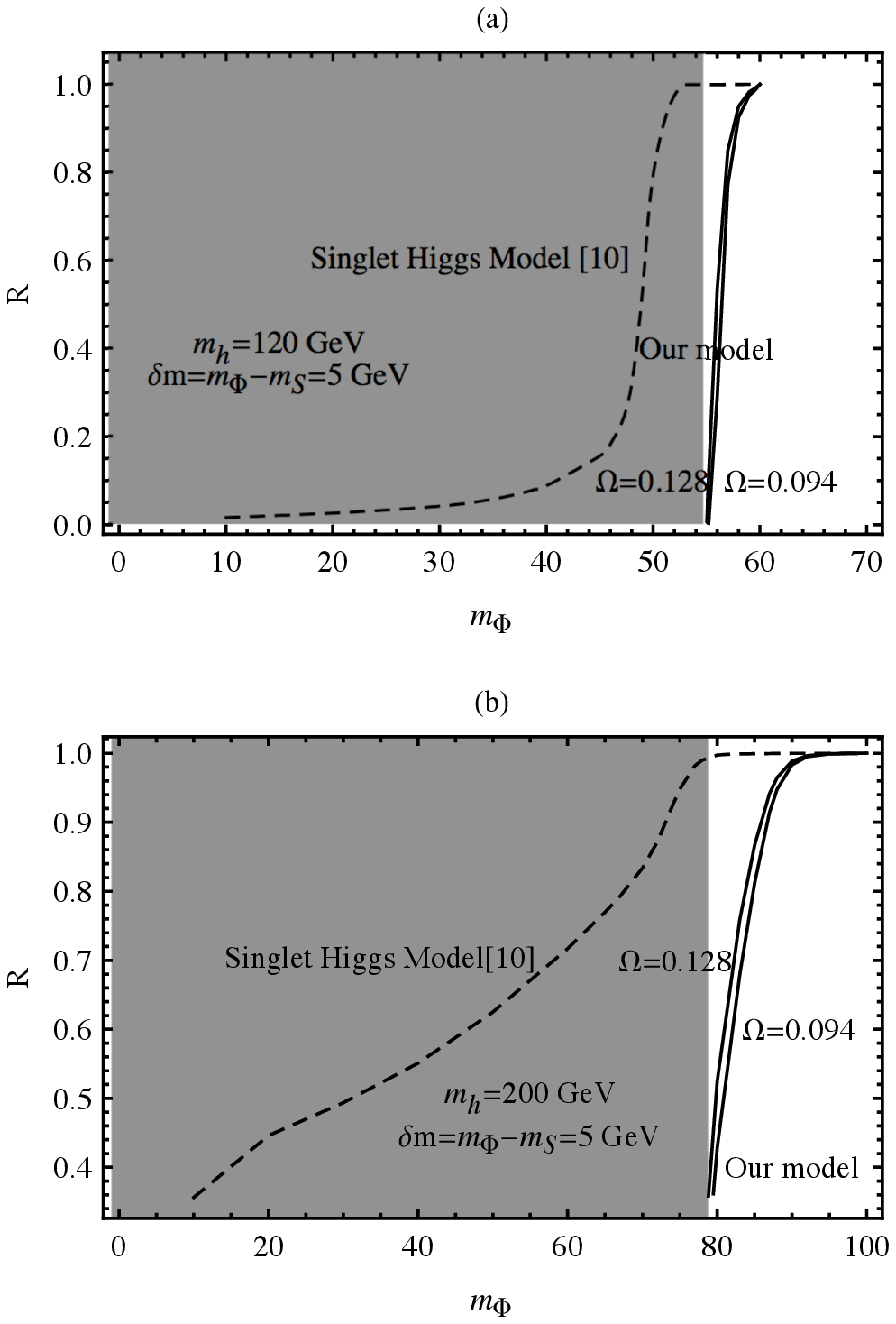, scale=0.7}}\cp{Plots of the ratio R as a function of $m_{\Phi}$ for our model with
$\Omega_\Phi h^2 = 0.128$ and $0.094$, (solid lines) when (a) $m_h=120$ GeV and (b) 200 GeV,
and for the model with a self-interacting dark matter \cite{sh} with $\Omega_\Phi h^2 = 0.128$ (dashed line).
Here the shadowed region represents the forbidden region by Fig. 4, which applies only to  our model.}
\ef

(a) {\it Case for singlet neutrino dark matter:}
In Fig. 8, we plot the value of $R$ as a function of $m_{\Phi}$ for (a) $m_h=120$ GeV  and (b) $m_h=200$ GeV, respectively.
Here, we fixed the value of $\delta m$ to be 5 GeV, and used the relation between
$\lambda$ and $m_{\Phi}$ which is obtained from the relic abundance constraints as explained before.
On calculating the decay rate of Higgs particle, we have used CalcHEP 2.4.5 \cite{calcHEP}.
As can be seen in Fig. 8,
the prediction for the value of $R$ is totally different from that in \cite{sh}.
But, in this case, we can probe the invisible Higgs decay by using $R$ only for the very narrow region
$55~(79)~ \mbox{GeV}< m_{\Phi}< 60~(100)~ \mbox{GeV}$ when $m_h=120~(200)$ GeV. This is because the lower limit of
$m_{S}$ has been determined to be $50 ~(74)$ GeV  for $m_h=120~(200)$ GeV, as shown in Fig. 4, and upper limit of $m_{\Phi}$
is constrained by the condition  $2m_{\Phi} < m_h$.
\fig [t]
\ct{\ep{figure=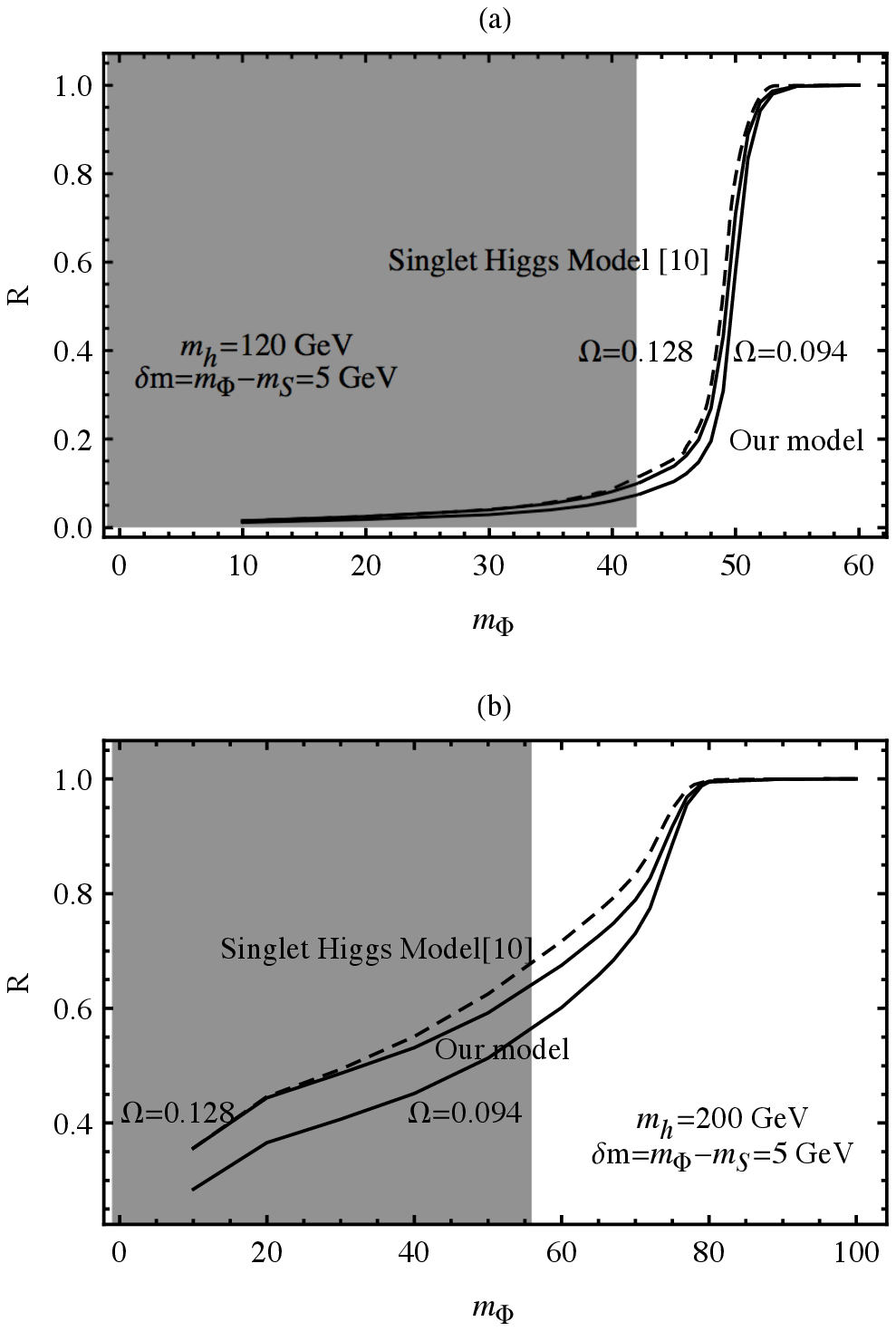, scale=0.7}}\cp{Plots of the ratio R as a function of $m_{\Phi}$ for our model with
$\Omega_\Phi h^2 = 0.128$ and $0.094$, (solid lines) when (a) $m_h=120$ GeV and (b) 200 GeV,
and for the model with a self-interacting dark matter \cite{sh} with $\Omega_\Phi h^2 = 0.128$ (dashed line).
Here the shadowed region represents the forbidden region by XENON10 Dark Matter Experiment \cite{xenon},
which applies only to  our model.}
\ef
{}From Fig. 8, similar to \cite{sh}, it turns out that the invisible Higgs decay width dominates the total width
everywhere except in the vicinity $2m_\phi =m_H$.
This means that a tremendous suppression of the observable Higgs signal may happen at the LHC.

(b) {\it Case for singlet scalar dark matter:} For this case, we plot the prediction for $R$ in Fig. 9
and the result is turned out to be almost the same as that in \cite{sh}. Similar to the above case,
the invisible width also dominates the total width everywhere except the region $2m_{\Phi} \simeq m_h$ \cite{sh}.
We notice that when  $m_h=120~ (200)$ GeV  and $\Omega h^2 =0.128$ the parameter space below $m_{\Phi}=42 ~(56)$ GeV are excluded by
the current bound obtained from XENON10 Dark Matter experiment.
In this case, if we impose the current bound from XENON10 Dark Matter Experiment \cite{xenon},
the mass of scalar boson $m_{\Phi}$ should be larger than 42 (56) GeV for $m_h=120~ (200)$ GeV and $\Omega h^2 =0.128$.
Then, the values of $R$ in our model is constrained to be
$0.1 - 1$ ($0.64-1$) where the upper limit is determined by the condition $2m_{\Phi}<m_h$ as it should be.\\

{\bf (ii) Case for $2m_{\Phi}>m_h$ }: \\
In this case, the singlet scalar bosons can be produced only through virtual Higgs exchange.
Similar to the previous case, the produced singlet scalar particles can be detected
as missing energy above an energy threshold, $E\geq 2m_{\Phi}$.
In this case, LHC is unlikely place for discovery of a missing energy signal, whereas future
linear collider might be a good place for detecting such a signal.

\section{Conclusion}

We have considered a variant of seesaw mechanism by introducing extra singlet neutrinos  and
singlet scalar boson and showed how low scale leptogenesis is realized in our scenario.
We have examined if the newly introduced neutral particles, either singlet Majorana neutrinos or singlet scalar
boson, can be a candidate for dark matter.
We have shown that the coannihilation process between dark matter candidate and the next lightest $Z_2$
odd particle plays a crucial role in generating the right amount of the relic density of dark matter  candidate.
We have also discussed the implications of the dark matter detection through the scattering off the nucleus
of the detecting material on our scenarios for dark matter candidate. {}From the recent result of XENON10 Dark Matter experiment,
we could get some constraint on the mass of singlet scalar boson.
In addition, we have studied the implications for the search of invisible Higgs decay at
LHC which may serve as a probe of our scenarios for dark matter.
\\

\noindent {\bf Acknowledgement:}
SKK is supported  by the KRF Grant funded by the
Korean Government(MOEHRD) (KRF-2006-003-C00069).
CSK is supported in part by CHEP-SRC and in part
by the Korea Research Foundation Grant funded by the Korean Government (MOEHRD)
No. KRF-2005-070-C00030.
\\


\end{document}